\documentclass[printer]{aa}
\usepackage{graphicx}
\usepackage{natbib}
\newcommand{\be}{\begin{eqnarray}}
\newcommand{\ee}{\end{eqnarray}}

\newcommand {\nbodypp}{\textsc{\mbox{nbody6\raise.4ex\hbox{\tiny++}}}}

\newcommand {\Msun} {\mbox{M$_{\odot}$}}
\newcommand {\Rsun} {\mbox{R$_{\odot}$}}

\begin{document}

\title{Encounter-driven accretion in young stellar clusters -  A connection to FUors?}
\author{S.~Pfalzner} 
\institute{I. Physikalisches Institut, University of Cologne, Z\"ulpicher Str. 77, 50937 Cologne, Germany}
\date{}

\abstract
{The brightness of FUors increases by several magnitudes within one to several years. The currently favoured explanation 
for this brightness boost is that of dramatically rising accretion from the disc material around a young star. The mechanism 
leading to this accretion increase is a point of debate.  }
{Choosing the Orion Nebula Cluster as representative we simulate accretion bursts driven by encounters in dense 
stellar environments. We investigate whether properties like rise and decay times, event frequency etc. speak 
for encounters as a possible cause for FUor phenomena. }
{We combine cluster simulations performed with the Nbody6++ code with particle simulations that describe the 
effect of a fly-by on the disc around a young star to determine the induced mass accretion. }
{The induced accretion rates, the overall temporal accretion profile, decay time and possibly the binarity rate 
we obtain for encounter-induced accretion agree very well with observations of FUors. However, the 
rise-time of one year observed in some FUors is difficult to achieve in our simulations unless the matter is stored 
somewhere close to the star and then released after a certain mass limit is 
transgressed. The severest argument against the FUors phenomenon being caused by encounters
is that most FUors are found in environments of low stellar density. We extend the discussion to eccentric 
binaries and gravitationally unstable discs and find that both models have similar
problems in achieving the necessary rise-times. }
{We find no conclusive answer as to whether the observed FUors are triggered by encounters. However, it seems 
there should exist an intense accretion burst phase - possibly an FU phase - early on in the development of 
dense clusters. We predict that in dense young clusters these outbursts should predominantly happen close to the cluster center and 
with large mass ratios between the involved stars. }

\titlerunning{Encounters - Connection to FUors?}
\authorrunning{Pfalzner}

\keywords{clusters - protoplanetary discs - circumstellar matter - ONC}
\maketitle

\section{Introduction}

For all stars, with the possible exception of massive stars, accretion of circumstellar 
matter is the most important formation mechanism. In a previous paper 
\cite[hereafter abbreviated as PTS08]{pfalzner:08} it was demonstrated that in a 
young, dense cluster environment, interactions with other cluster members lead to accretion 
bursts where up to 6-7\% of the disc material around the young star can be accreted within a short time (10$^2$-10$^4$ yrs). 
The induced accretion rates can be as high as those observed in FU Orionis objects 
(FUors), raising the question of whether FUors could result from such 
encounter events.  

FUors are a small class  of pre-main-sequence objects named after
FU Orionis - the first observed object of this type. FUors show large brightness increases 
(5-6 mag) on short timescales (0.5 - 15 years) coupled with significant changes in spectral type.
Apart from these classical FUors, about 20 FUor-like objects  are currently known 
\citep{vittone:05}: these have spectral similarities to classical FUors, but for which no 
eruption has been witnessed \cite{reipurth:02}. 

The mechanism by which these eruptions occur is still poorly understood 
\cite{herbig:03,reipurth:04,hartmann:04}. 
The most widely accepted view is that the outbursts are caused by some type of accretion 
disc instability. In this framework the outburst is not produced by the star itself 
but results from an increase in the surface brightness of the circumstellar disc 
surrounding the pre-main sequence stars \citep{hartmann:85,hartmann:96}. 
It should be mentioned that Herbig, Petrov \& Duemmler (2003) suggest an alternative
model in which the outburst occurs in an unstable young star rotating near breakup velocity. 
The various accretion disc models \cite{clarke:mnras90,bell:94,kley:99,lodato:2004,vorobyov:05,mejia:05}
offer different mechanisms for triggering the accretion outburst, such as thermal and 
gravitational instabilities possibly combined with magneto-rotational instabilities \cite{armitage:01}.

In the following we summarize the properties of the different proposed instability-triggering mechanisms:

(i) The thermal instability model \cite{bell:94} considers  a disk fed at high mass accretion rate 
from a surrounding envelope, which becomes unstable at the inner edge. The instability 
propagates outwards, producing a slowly rising luminosity compatible with the light curve of V1515 Cyg.
However, this model cannot explain the rapid rise-time of FU Ori and V1057 Cyg \cite{vittone:05}. 
%As the instability must be first triggered at a large radius, the instability propagates outside in. 
Another problem is the size of the outburst region, where the thermal instability model predicts an 
outer radius of $\sim$ 20 \Rsun.
Constructing a radiative transfer disc model of FU Orionis, Zhu et al. (2007) find that the outer radius of the 
accreting region is 0.5 - 1.0 AU and conclude that thermal instability models have difficulties explaining 
this fact. 

(ii) Gravitational instabilities occur only if the disc masses involved are large enough. According to 
Vorobyov \& Basu (2006) the accretion bursts are associated with the formation of dense protostellar/protoplanetary
embryos, which are later driven onto the protostar by the gravitational torques that develop in the disk.
%The burst phenomenon is quite sensitive to the amount of angular momentum in the cloud, and the bursts are 
%more frequent and intense for greater values of the ratio of the magnitudes of rotational and gravitational 
%energy. 
There is some ambiguity about the duration of the outburst and the rise-times.
Armitage et al. (2001) see an outburst duration of $\sim$ 10$^4$ yr, two orders of magnitude longer
than the observed ones, whereas Vorobyov and Basu (2006) claim  very short ($<$100 yr) vigorous
($\stackrel{.}M =  1- 10 \times 10^{-4}$ \Msun/yr) accretion bursts, which are intervened
by longer periods ( $>$ 10$^3$ yr) of quiescent accretion.
In their model the frequency of the bursts decreases noticeably with time, and no bursts are seen
after t $\sim0.3$ Myr. 
The high disk mass required by this model ( ${\approx}1~M_{\odot}$) is consistent with 
measurements of some FU Orionis objects \cite{henning:98,sandell:01}, but 
the measured disk mass of V1057 Cyg ( ${\approx} 0.1~M_{\odot}$, Sandell \& Weintraub 2001) seems to 
be too low to cause any observable effect.

(iii) A related idea is the passage of a companion star close to the primary star causing 
an increase in accretion which was originally proposed by Toomre \cite{kenyon:00}.
This concept was investigated by Bonnell \& Bastian (1992b) for the case of wide eccentric binaries.
Clarke \& Syer (1996) studied the situation where a protoplanetary/protostellar companion close
to the primary causes a gravitational instability. 

Here we investigate the properties of the very similar situation to the latter - encounters in young dense 
clusters  as described in PTS08. The difference in this approach is that whereas
Bonnell \& Bastian (1992) and Clarke \& Syer (1996) look for specific situations which would produce the
accretion rates necessary for FUor, we start from the induced accretion processes that  
inevitably occur in young clusters and ask whether these would be interpreted as FUors.  

Before we discuss the likelihood of the passage of a star causing FUor outbursts 
we first summarize the known FUors properties.

\section{Properties of FUors and FUor-like stars}

The most striking  feature of FUors is, as already mentioned, the increase in brightness of 5-6 mag.
While the rise-times for outbursts are usually very short ($\sim$ 1-10 yr), the decay
timescales range from decades to centuries. 

FUors are spatially and kinematically associated with star-forming regions 
\cite{herbig:77,hartmann:85}. 
Other observational properties of FUors are: optical spectrum resembling F-or G-type supergiant
stars \cite{herbig:77}; near-infrared spectra similar to K-or M-type cooler giant stars \cite{mould:78};
they are often associated with bright reflection nebulae \cite{goodrich:87}; many are heavily 
extincted; and they mostly have large infrared excess emission from circumstellar dust 
(Weintraub et al 1991). Some FUors have collimated Herbig-Haro (HH) jets and wide angle molecular 
outflows. These properties suggest that at least some of the FUors are quite young stellar objects. 

Another characteristic is that they show double peaked absorption-line profiles
\cite{hartmann:96}. In the most widely accepted model \cite{kenyon:93} these peculiarities 
are explained by a combination of an accretion disc and an envelope. In contrast
to the discs of most classical T Tauri stars, which are primarily heated by radiation from the central star 
(e.g., Kenyon\& Hartmann 1996), the disc heating of FUors is dominated by local accretion 
energy release at least in the inner disk regions.

%Evidence for large, massive disklike structures capable of replenishing the 
%smaller accretion disks (Weintraub et al. 1991) supported the idea that FUor 
%outbursts primarily occur in an early phase encompassing the first few hundred 
%thousand years of protostellar evolution.

\begin{table} 
\caption{Properties of FUor and FUor-like objects}\label{tab:prop} 
\begin{tabular}{ll} 
\hline
\hline 
Property         &                                     \\
\hline \\

accretion rate      & 10$^{-7}$-10$^{-4}$ \Msun/yr\\
stellar mass        & $\sim$ solar-mass or lower\\
rise-times          & 1-10 yr\\
decay times         & 10 - several 100 yr\\
spectral type       & large infra-red excess\\
absorption lines    & double peaked\\
location            & star-forming regions\\
                    & rarely in dense clusters\\
age                 & probably younger than T-Tauri stars\\
                    & but older systems exist, too \\
disc size           & often larger than T-Tauri stars\\
disc mass           & often more massive than T-Tauri stars\\
                    & but low disc masses exist, too\\
binarity rate       & high \\
\hline 
\end{tabular} 
\end{table}

Some FUors have high disc masses (Henning et al. 1998; Sandell \& Weintraub 2001),
typically a few tenths of a solar mass with a few FUor discs being even more massive. 
These discs are more massive than those of Class II T Tauri stars,
which usually have an order of magnitude lower mass ($\sim$ 0.01 \Msun\, Osterloh \& Beckwith 1995),
resembling more the disc masses of Class I objects \cite{sandell:01} ---
another indication that some FUors likely are extremely youthful compared to T Tauri stars.
However, there exist exceptions as well with much lower disc masses undergoing FU outburst 
like V1057 Cyg ( ${\approx} 0.1~M_{\odot}$, Sandell \& Weintraub 2001) and FU Orionis itself ($<$ 0.02 \Msun).
AR 6 and  FU Orionis are Class II objects as they lack strong sub-mm fluxes,
indicating that their circumstellar envelopes and disks are relatively depleted. The implication
therefore is that FUor eruptions can occur during any phase of the stellar evolution.

In addition, the limited number of measured disc sizes of FUor discs also hints at FUor discs being  
larger than those of T Tauri stars \cite{weintraub:91,sandell:01}, more like Class 0 and Class I objects  
\cite{chini:97,huard:99,hogerheijde:00,sandell:01},
or possibly even larger than those \cite{sandell:01}. 
This might suggest that FUor events are more likely among sources surrounded by large disks. 

A growing number of FU Orionis objects are found to be binaries \cite{reipurth:04}
or even part of a multiple system. 
For example, the FU Orionis object Z CMa is a binary system \cite{koresko:91}, 
and  a companion to FU Ori has been found  \cite{reipurth:04,wang:04}. 
However, it has not yet been proved that all FU Orionis objects have companions. 

Most FUors are found in dense regions of dark clouds and quite often are 
heavily reddened by local extinction \cite{sandell:01}. Few FUors
or stars with FUor-like spectra are found in regions of clustered star formation such as 
Tau-Aur, ρ Oph, NGC 1333, IC 346, or the Orion Nebula Cluster. Instead, they are mostly found
in regions of low star-formation activity. An exception is the double system AR 6A and 6B which is 
part of the Spokes young stellar cluster \cite{moriaty:08}.

%The observations suggest that the typical low-mass star accretes  10\% of its mass in 
%“FU Ori” accretion (by which we mean high accretion rate states, not necessarily outbursts)
%\cite{hartmann:96}

The common properties of FUors and FUor-like objects are summarized in Table \ref{tab:prop}.
However, the statistics are still poor and these correlations must be confirmed through 
additional studies; an obvious statement, if one considers the known properties of individual 
FUor sources listed in Table \ref{tab:fuorcat} in the Appendix.

%Broad cross-correlation peaks indicate rotational velocities v sin i $\sim$
%100 km s$^{-1}$, much larger than measured for T Tauri stars \cite{greene:08}.

\section{Method}

We performed two types of simulations: tree code simulations to study the 
effect of encounters on star-disc systems \cite[for details see ][]{pfalzner:aa05} and 
cluster dynamics simulations \cite[see][]{olczak:apj06} performed with the 
Nbody6++ code \cite{spurzem:mnras02} to determine the likelihood and degree of 
encounter-triggered accretion.

In the  cluster simulations, for simplicity all stars were assumed to be initially single.  
Cluster models were set up with a spherical density distribution $\rho(r)\propto r^{-2}$ and a Maxwell-Boltzmann 
velocity distribution. The stellar masses were generated randomly according to the mass function 
given by Kroupa (2002) in a range $50 \Msun \ge M^* \ge 0.08 \Msun$. 
During the simulation information of all perturbing events on each stellar disc was recorded. 
The quality of the dynamical models, which we chose to be in virial equilibrium, was determined 
by comparison to the observational data of the ONC \cite{mccaughrean:msngr02}.

\begin{figure}[t]
\resizebox{\hsize}{!}{\includegraphics[angle=-90]{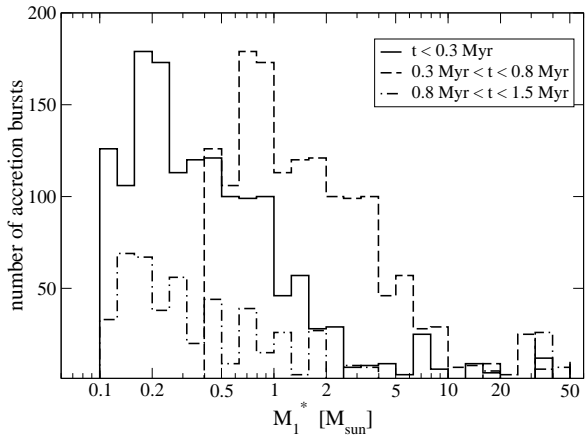}}
\caption{The number of encounters that lead to accretion bursts as a function of the primary
mass $M_1^*$ for different time intervals of the cluster development.}
\label{fig:enc_mass}
\end{figure}

The encounter simulation part of the investigation is mainly based on our earlier work \cite{pfalzner:08}, 
which includes a parameter study of the situation where the disc of a star of mass $M_1^*$ 
is perturbed by the flyby of a second star of mass $M_2^*$. In these simulations we assumed disc 
parameters typical of T-Tauri stars : the discs to be of low-mass (0.01 \Msun), the disc surface density 
to have a  1/r-dependence initially and the disc size to scale with the stellar mass as 
$ r_{\rm {d}} = 150AU \sqrt{M_1^*[\Msun]}$.

The code applied here cannot describe the disc dynamics simultaneously with the accretion process 
as the mass density - and so the required number of simulation particles - would be very high
close to the star. More importantly, the required temporal resolution for the accretion process
would be so high that the overall disc dynamics could not be monitored for 
long enough times.  
The main question is whether perturbations to the inner disc area by an encounter can induce accretion.
Whether this accretion happens directly by a particle stream or only some related
mechanism is of secondary importance. 
Therefore we resort to the approach of Bonnell \& Bastien (1992) and Vorobyov \& Basu (2006), 
who modelled accretion in FU Orionis objects by investigating the amount of 
matter reaching a sphere of a certain size around the central star - here a region of radius 1AU.
So the word ``accretion'' is used in this sense hereafter.
We model only gravitational interactions, pressure and viscous forces have been neglected. 
In principle the latter could hinder matter reaching the inner disc areas due to pressure 
gradients. However, here the velocity of the instreaming matter is so
high that this is in most cases not a problem.

%It is very difficult wether this approach over- or underestimates the actual
%accretion rate. On the one hand matter reaching the 1AU region often is on 
%highly eccentric orbits, so when it reaches the accretion radius it is close to
%its periastron and its velocity is sometimes very high. Therefore it might be
%difficult to accrete. On the other hand, it is commonly believed that viscous 
%processes are able to recirculize the highly excentric orbits
%of the particles after the encounter after some time and ease accretion that way.

\begin{figure*}
\resizebox{\hsize}{!}{\includegraphics[angle=0]{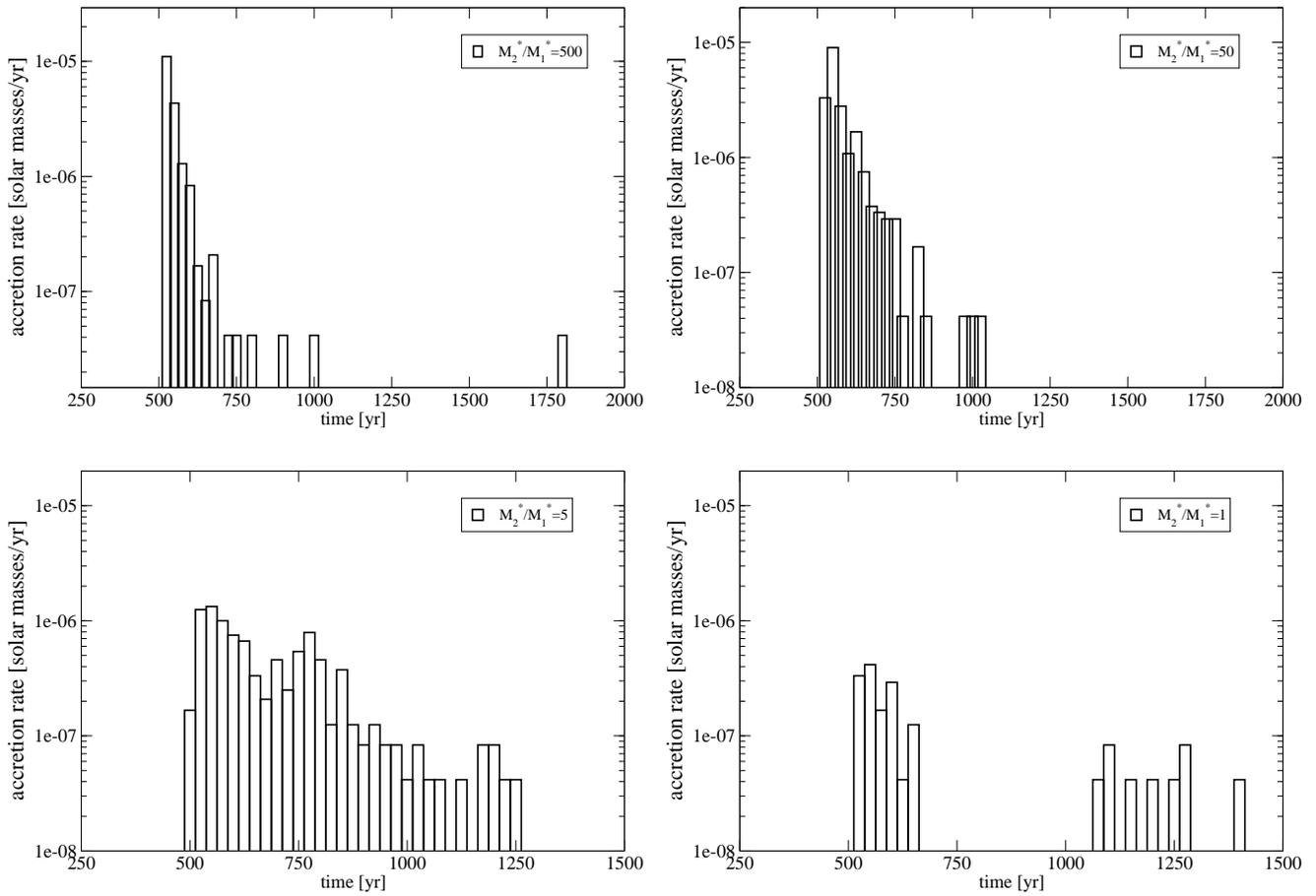}}
\caption{Accretion rate as a function of time for an parabolic encounter with
$M^*_2/M^*_1$ = 500, 50, 5, 1 and a perihel distance of 1.5 $r_d$ assuming $m_d$ = 0.001 \Msun.}
\label{fig:accrate_time_diff}
\end{figure*}

\section{Connection between encounter-triggered accretion and FUors?}

In the following we present our simulation results of encounter-triggered accretion bursts and compare them with the characteristics of FUors
listed in Table \ref{tab:prop}. 

Our simulations have already shown that a fly-by leads to a very rapid increase in accretion. We demonstrated in
PTS08 that the maximum amount of matter accreted in a single prograde, coplanar encounter event is 
up to 6-7\% of the disc mass. For typical disc parameters of a solar-mass star T Tauri star
($m_d \sim$ 0.01 \Msun, $r_d \sim$ 150 AU) this would be equivalent to an encounter-averaged 
accretion rate of 10$^{-7}$-10$^{-6}$ \Msun/yr. In the initial phases the instantaneous accretion 
rates can be much higher reaching  10$^{-4}$ \Msun/yr, which is what one observes in FUors.
Submillimeter observations show that the disk masses in FUor systems 
($m_d^{FUor}\sim $ a few times 0.1 \Msun\, \citep{sandell:01}) 
are much higher than those of T Tauri stars ($m_d^{TT}\sim 0.01 M\sun$ \cite{osterloh:95}). 
Such massive discs guarantee similar high accretion rates even in encounters that are less favourable
for accretion like non-coplanar or more distant encounters etc.  

\subsection{Low-mass stars}

As demonstrated in PTS08, in an ONC-like environment each individual massive star is much more likely to 
undergo  accretion outbursts due to encounters than any lower mass stars. At first sight
this seems to be in contrast to the fact that the observed FUors are mostly low-mass stars 
($\sim$ 0.1 - 2 \Msun). However, as there are much more low-mass 
stars than massive stars in the cluster, at any given time the number of outbursts of all low-mass stars is
higher than that of all massive stars (see Fig.~3 in PTS08). Therefore one is far more likely
to observe an accretion burst in a low-mass star than in a massive star, which does agree with the
observations in FUors.

\subsection{Rise and decay times}

\begin{figure}
\resizebox{\hsize}{!}{\includegraphics[angle=0]{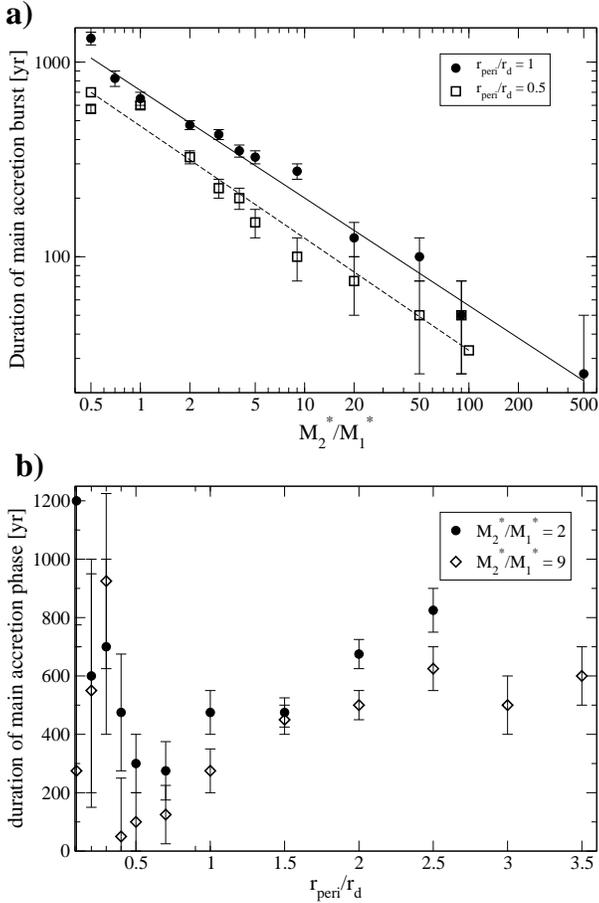}}
\caption{Time required until the star has accreted 95\% of the mass, it eventually accretes from its own disc
as a function of a) the mass ratio
$M_2^*/M_1^*$ and b) the relative periastron $r_{peri}/r_d$ in a single encounter. The lines serve to
guide the eye.}
\label{fig:accrate_duration}
\end{figure}

With shorter rise than decay times the overall shape of the temporal development of the accretion induced 
by encounters (see Fig.~\ref{fig:accrate_time_diff}) appears to be similar to that observed in FUors.  
Our simulations show that the accretion induced by strong encounters characteristically rises within 
10 to several 10s of years, afterwards the accretion declines typically on timescales of several 10s 
to several 100s of years. 

The accretion rise and decay times depends on the 
interaction strength, with stronger interactions leading to shorter but more intense
accretion outbursts. The reason is that the
time where both stars are close together is shorter for stronger interactions.
Fig.~\ref{fig:accrate_duration} shows the outburst duration (here defined as the time in which 95\% of the total accreted matter is
accreted) as a function of the mass ratio $M_2^*/M_1^*$ and the periastron distance in the encounter.
It can be seen that the mass ratio has to be large in order to obtain strong outbursts of short
duration.
%as illustrated by Fig.~\ref{fig:accrate_duration}. 

Although there exist encounters that produce strong short outbursts, we never observe rise-times 
on the scales of 1 year in our simulations, as observed for FU Ori. 
However, this does not necessarily mean that FUor events are not triggered by 
encounters. The crude definition and modelling of the accretion process in our simulations 
could cause the accretion times to appear longer than they actually are: 
As  Hartmann \& Kenyon (1996) noticed, the interaction must perturb the disk at 1-10 AU to 
produce sufficiently short rise 
times, because perturbations  at larger distances naturally show much longer
(viscous) rise-times. However, in our simulations the inner edge of our disc is at 10AU, so that we 
do not resolve the area down to 1AU as required. Moreover, the pressure and viscous
forces neglected here could also alter the rise and decay times.

\subsection{Sparcity in cluster environments}

\begin{figure}
\resizebox{\hsize}{!}{\includegraphics[angle=-90]{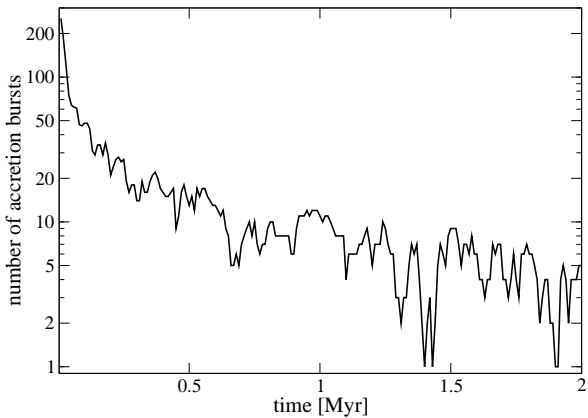}}
\caption{The number of encounters in any 10$^4$ yrs long time-interval that lead to accretion bursts as a function of the cluster age in an ONC-like cluster.}
\label{fig:enc_mass2}
\end{figure}

The strongest argument against FUor outbursts being triggered by encounters is that few 
FUors or stars with FUor-like spectra are found in young clusters such as Tau-Aur, ρ Oph, 
NGC 1333, IC 346, or the Orion Nebula Cluster. If encounters  triggered FUor outbursts
it is in these dense environments that one would expect most of the bursts to happen. 
However, FUors are mostly found in regions of low stellar density \cite{herbig:03}. An exception is the double 
system AR 6A and 6B which is part of the Spokes young stellar cluster \cite{moriaty:08}.
This implies that either FUors are much less frequent  in high-density regions or else have 
fewer outbursts.

Greene et al. (2008) suggested that  interactions between stars in clusters could disrupt disks, 
perhaps eliminating much FUor activity during the Class 0 or Class I protostellar evolutionary phases.
This would agree with the fact that encounter-triggered accretion bursts are a strong function of time 
in dense environments like the
ONC. Fig.~\ref{fig:enc_mass2} shows the number of encounter-triggered accretion-bursts in any 10$^4$ yr
time-interval as function of cluster age for an ONC-like cluster. It can be seen that there are a 
factor 10-20 fewer accretion outbursts nowadays ($\sim$ 1Myr) than in the early phases of the cluster
development. The reasons are that the cluster becomes less dense and the discs lose mass due to
tidal interactions.

\section{Encounter or Binary?}

Many of the observed FU Orionis objects are classified as binaries 
\citep{koresko:91,kenyon:93,rodrigues:95,hartmann:96,aspin:03} and 
some of the companions are surrounded by a disc themselves \citep{reipurth:04,quanz:06}.
The distances of the companions are relatively large ($\sim$ 40 - 5000 AU), so they
are all visual binaries. Since usually at least one of the stellar masses is unknown,
in principle the two stars could actually undergo an encounter rather than being a binary.
What speaks against this is again the low stellar density in most of these regions
where FUors are observed.

However, the accretion burst induced by an encounter and by a binary that recently formed
would be nearly indistinguishable anyway, as the binary would have to have a high eccentricty to
display the accretion characteristics of a FUor. 
%The collapse and fragmentation of elongated clouds 
%has been shown to form binary systems with large eccentricities  \cite{bonnell:92b}.
The binary mechanism is attractive, since most stars are members of multiple systems.

%One argument against binaries as cause of the FUor phenomenon was pointed out by
%Hartmann \& Kenyon (1996). They argued that since the interaction must strongly perturb the disk 
%at 1–--10 AU to produce short rise-times, it would be necessary to assume 
%that the orbital distribution among binaries was originally much more concentrated at small 
%perihelion distances than presently observed in the field. 
%However,  our results suggest that large mass ratios could lead to the required 
%perturbation, so that the requirements on the orbital distribution are less stringent 
%(see Fig.~\ref{fig:accrate_duration}). 

The interaction must strongly perturb the disk at 1–--10 AU to produce the short rise-times
of FUors. Often this requirement is translated as the necessity of small periastron distances
in the binary. Our results suggest that large mass ratios could equally lead to the necessary 
perturbation, so that the requirements on the orbital distribution are less stringent 
(see Fig.~\ref{fig:accrate_duration}).

\subsection{Mass from primary or secondary}

\begin{figure}
\resizebox{\hsize}{!}{\includegraphics[angle=0]{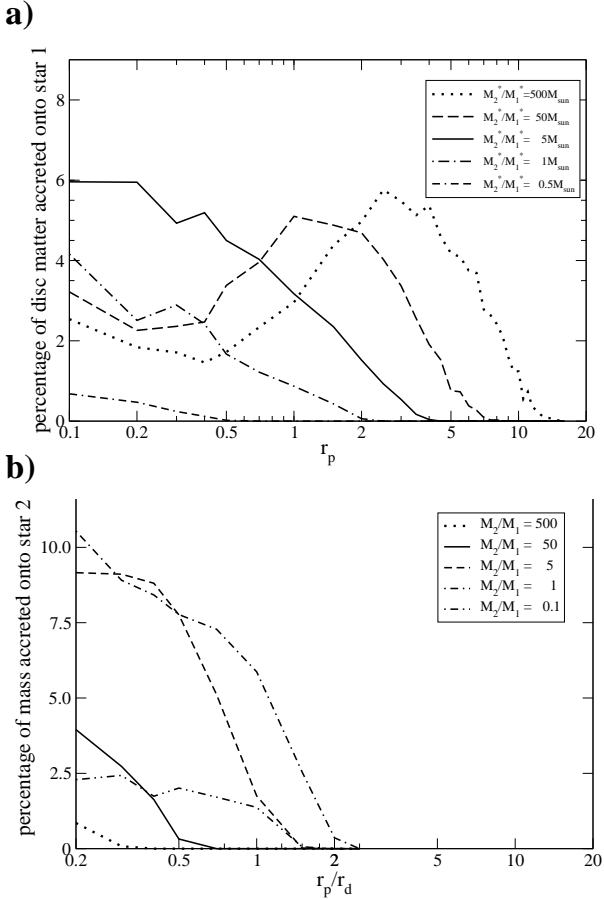}}
\caption{Percentage of disc mass accreted onto star 2 as 
a) function of the relative periastron distance $r_p/r_d$ and b)
function of relative mass ratio $M_2^*/M_1^*$ in parabolic, prograde 
coplanar encounters.}
\label{fig:acc2}
\end{figure}

Bonnell and Bastien (1992a) studied how wide eccentric binaries could trigger FUor outbursts. 
As the parameter space is rather extensive they studied only a few selected configurations. 
%For the investigated cases 
They found that for unequal-mass binaries the more massive component has a stronger tendency 
to be obscured by the surrounding medium, has the more massive disc, induces a larger accretion 
rate and that accretion will be much larger onto the lower mass secondary than onto the primary.
They suggested that it is actually this sudden increase of matter in the 
secondary's disk obtained from the primary's disk that could describe the FU phenomenon. 
%They argue that  due to the comparable orbital velocity of the outer parts of the primary's disk at 
%closest approach, 
Because the accreted matter obtained from the primary's disk has little rotational velocity in the 
secondary's reference frame, it can be deposited at arbitarily small radii or accreted directly.
%Although this would lead to enhanced accretion in the infrared in both discs, the
%accretion onto the secondary would be mainly visible in the optical. 
Mass would fall predominantly onto the center of mass close to the primary and would form a disc which 
obscures the primary. By contrast, the secondary would accrete directly with less extinction, so that 
the large far-infrared excess could be explained by the presence of a IR companion.

Reipurth \& Aspin (2004) proposed that such events may be a natural consequence of interactions between
two components that spiral in toward each other following the breakup of an unstable triple system. 
Alternatively, Greene et al. (2008) suggested that a dense cluster environment might accelerate the 
orbital evolution, again causing FUor eruptions in clusters to occur mostly during the Class 0
or Class I stages (Reipurth \& Aspin 2004). 

\begin{figure}
\resizebox{\hsize}{!}{\includegraphics[angle=-90]{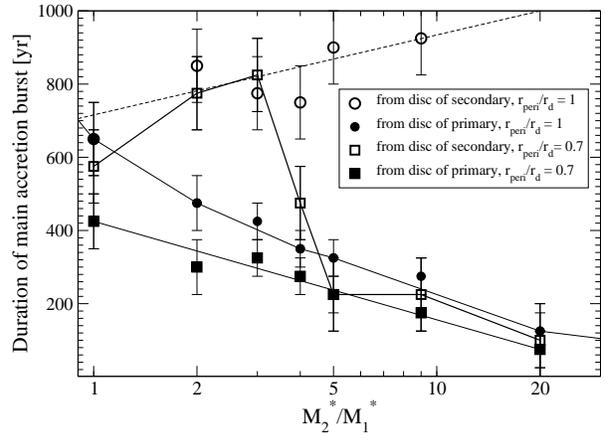}}
\caption{Comparison between the time required until the star has accreted 95\% of the mass, 
it eventually accretes from its own disc (full circles) and the mass it accretes from the disc 
of the secondary star as a function of the mass ratio $M_2^*/M_1^*$.}
\label{fig:accrate_time}
\end{figure}

Although the approach of an accretion burst in the secondary is appealing, there are some problems.
First, accretion of matter onto the secondary can only happen for relatively close encounters,
$r_{peri}/r_d < $ 2 (see Fig.~\ref{fig:acc2}). However,  some of the discs observed around FUors 
are relatively large, so this might be not so stringent. Second, the overall accretion time of the matter 
onto the secondary is longer than that onto the primary (see Fig.~\ref{fig:accrate_time}) and,
more importantly, the rise-time of the accretion burst is longer, too. Only if $M_2^*/M_1^* \gg$ 1,
can the rise-time become $\sim$ 10 yrs, but in this case the accretion rate onto the secondary declines
considerably. As the rise-time is the most critical point anyway, there is a problem here.

However, the two sources of accretion - onto the primary and onto the secondary - could
potentially hold the key to distinguishing between binary/encounter driven outbursts and
instabilities caused by gravitational unstable discs or planetary companions. The first do
have these two sources of accretion, whereas the latter do not. Our simulations show that
relatively independently of the actual encounter parameters, the accretion onto the secondary
always happens $\sim$ 100 yrs before that onto the primary. This possibly means that 100 yrs before
the outburst in the visual there takes place an outburst in the infrared. If this
would be observed this would give a clear distinction between the two groups of triggering
mechanisms.

\section{Conclusion}

We have shown that there are many properties that encounter-induced accretion bursts 
%happen inevitably in young dense clusters like the ONC 
and FUors events have in common. The accretion rates, the accretion profile and 
the overall burst duration in our encounter simulations are basically identical to those
observed in FUors. From our simulations it follows that encounter-induced accretion outburst 
would be mainly visible for low-mass stars and would currently not be distinguishable from visual binary. 
Both properties are in accordance 
with the fact that many FUors are low-mass binaries. However, there are two problems with FUors being
induced by encounters - (i) FUors predominantly (but not exclusively) are observed in 
regions of low stellar density and (ii) the rise-times in encounter-induced accretion bursts
seem to be longer than those observed for FUors.

%However, our investigations indicate that other models share this latter problem. 
%All descriptions of gravitationally-triggered accretion bursts whether the trigger is an encounter, a
%binary, a massive disc or a planetary companion result in rise-times longer than $\sim$ 1 yr.
%This does not necessarily mean that none of these models can lead to FUor events, but probably
%on the contrary each of them could trigger FUor events but the current modelling of the actual accretion 
%process as such is to crudely resolved.

If the latter problem is real, then due to the similarity in mechanism other models of 
gravitationally-triggered accretion bursts whether the trigger is an encounter, a
binary, a massive disc or a planetary companion possibly share this problem of too long rise times.
However, this does not necessarily mean that none of these models can lead to FUor events:
on the contrary each of them could trigger FUor events and the long rise times might actually be an 
artefact of the crude accretion modelling (disc cut-off at 10 AU and exclusion of pressure/viscous 
forces).

The obvious question that arises is why are the spectra of FUors and FUor-like stars not more similar 
to the spectra of Class I protostars with large veiling and relatively high accretion
rates 10$^{-6}$ \Msun\,/yr? Greene et al. (2008) suggested that higher accretion rates are possibly 
needed to produce luminous disks with absorption features similar to those found in FUors.

Given the fact that the properties of FUors can differ considerably - from young stars with massive 
large discs in sparse regions to stars with low disc masses in cluster environments -, 
it may be too simplistic to assume there is only one way to trigger FUor outbursts. 
There may be actually several ways for FUors to occur - some externally triggered by
encounters, binaries, disintegration of a triple system and planetary companions
but as well others caused by diverse instabilities in disks around single stars.
FUor observations could then be just a selection effect requiring a certain accretion rate 
to see FUor spectra. This would explain as well the preference but not exclusiveness for FUor 
phenomena to occcur in young stars with high disc mass and large disc size. 

This interpretation only works as long as there are no FUors found with an accretion rate
less than 10$^{-6}$ \Msun\,/yr or stars with accretion rates higher than 10$^{-5}$ without FUor 
characteristics.

If this is the case then we would expect many FUors in dense clusters younger than 0.3 Myrs,
because here encounter-triggered accretion burst should be quite common. They would  mainly occur 
close to the cluster center and if the mass ratio of the encounter partners is high. If the encounter is close
enough an outburst in the infrared would precede one in the visible.

\section*{Acknowledgments}

We want to thank B. Reipurth for his very helpful comments.
Simulations were partly performed at the John von 
Neumann Institute for Computing, Research Centre J\"ulich, Project HKU14.

\bibliographystyle{apj}
%\begin{thebibliography}{alpha}

\onecolumn

\begin{table}%T1 %\centering \par 
\caption{Observed properties of FUor and FUor-like objects ($^1$ Malbet et al. 2005, 
          $^2$ Sandell \& Weintraub 2001,  $^3$ Reipurth et al. 2007,
          $^4$ Movsessian et al. 2006 $^5$ Kospal et al. 2008,         $^6$ Vittone \& Errico 2005, 
          $^7$ Rodriguez 1998,        $^8$ Quanz et al.  2006,           $^9$ Koresko 1991) 
          \label{tab:fuorcat}} 

\begin{tabular}{lccccccc} 
\hline
\hline 
Object       & t(Rise)      & t(Decay)       & Stellar Mass & Disc Mass$^2$ & Accretion rate           & Binary & Disc size \\
\hline 
BBW~76       &              & $\sim$ 40 yr   &              & 0.15 \Msun\,\                                                    \\
CB34~V       &              &                &              & 0.2  \Msun\,\                                                    \\ 
FU~Ori       & $\sim$ 1 yr  & $\sim$ 100 yr  & 0.35 \Msun\,\  & $<$ 0.02 \Msun\,\ & 6.5 $\times$ 10$^{-5}$ $^1$ & yes$^1$, k, 217 AU $^8$\\ 
V1057~Cyg    & $\sim$  1 yr & $\sim$  10 yr                                                                                   \\ 
V1515~Cyg$^{3}$& $\sim$ 20 yr & $\sim$  30 yr&              & 0.13 \Msun\,\ &                                                    \\
L1551 IRS 5   &             &                &              & 0.23 \Msun\,\ &                             & yes, 45 AU $^7$ \\ 
PP~13S        &             &                &              & 0.1 \Msun\,\  &                             &      & 2000AU$^2$            \\  
Re~50~N~IRS1  & \\ 
V1331~Cyg     & \\ 
V1735~Cyg     & $<$ 8 yr      & $>$ 20 yr        &               & 0.42 \Msun\,\ &\\
V346~Nor      & $<$ 5 yr      & $>$  5 yr        &               &             &                             & yes? \\
V883~Ori      &              &               &               & 0.39 \Msun\,\ &\\ 
Z~CMa         &              & $>$  100 yr    & 1.1 \Msun\,\ &  & 7.9 $\times$ 10$^{-5}$                       & yes,93 AU $^9$\\ 
RNO~1B        &              &               &               &             &        & yes, 5000 AU $^8$ \\ 
RNO~1C        &              &               &               &             &        & yes, 5000 AU $^8$ \\ 
AR~6A         &              &               &               &             &        & yes, 2240 AU $^8$ \\ 
AR~6B         &              &               &               &             &        & yes, 2240 AU $^8$ \\ 
OO~Ser        & \\ 
V1647~Ori     & \\ 
V733 Cep$^3$  & &  $>$ 20 yr & \\
RNO 127$^4$       & &  $>$  9 yr & \\
Parsamian 21$^5$  & & &  & $<$ 0.3 \Msun\,\ & 0.3 \Msun\,  \ & & 360 AU\\
\hline 
\end{tabular} 
\end{table}

%\begin{figure}
%\resizebox{\hsize}{!}{\includegraphics[angle=0]{acc_enc_time.eps}}
%\caption{Arretion rate as a function of time for an parabolic encounter with
%$M_2/M_1$=500,50,5,1 and a perihel distance of 1.5 $r_d$.}
%\label{fig:accrate_time}
%\end{figure}

\end{document}